\documentclass[format=acmsmall,authorversion,nonacm]{acmart}
\pdfoutput=1
\newenvironment{codelet}%
{\begin{center}\begin{minipage}{0.85\textwidth}}
{\end{minipage}\end{center}}
\newcommand{\code}[1]{\texttt{#1}}

\usepackage{fancyvrb}
\usepackage{csquotes}
\begin{document}

\title{Dynamic String Generation and C++-style Output in Fortran}

\author{Marcus Mohr}
\orcid{0000-0003-2942-8484}
\affiliation{\institution{Geophysics, Department of Earth and Environmental
    Sciences, Ludwig-Maximilians-Universit\"at M\"unchen}
  \city{Munich}
  \country{Germany}}
\email{marcus.mohr@lmu.de}

\begin{abstract}
Using standard components of modern Fortran we present a technique to
dynamically generate strings with as little coding overhead as possible
on the application side. Additionally we demonstrate how this can be extended
to allow for output generation with a C++ stream-like look and feel.
\end{abstract}

\keywords{Modern Fortran, Strings, Code Reduction, Stream-like Output}

\ccsdesc[500]{Software and its engineering~Polymorphism}

\maketitle

\citestyle{acmauthoryear}

\section{Motivation}
\label{sec:motivation}
Scientific simulation codes do not only perform large-scale I/O for reading
input datasets and storing final simulation results. Especially in the case
of long-running simulations, they also typically generate log messages to
inform the user on various details of the simulation run, ranging from
echoing steering parameters, over current time-step values, up to the progress
of iterative solvers and many more. These get send to either the
screen/terminal or a dedicated logfile or both.

In larger projects one quickly reaches a point where aspects such as the
following become highly desirable
\begin{itemize}
\item employ a 'nice' and uniform formatting for log messages
\item allow users/developpers to select different levels of verbosity
  (e.g.~debug, info, warning)
\item have different levels of indentation or another way to signal which
  program component logged a message 
\item allow switching the message destination (terminal, file or both)
\item \ldots
\end{itemize}
Enforcing formatting rules and checking e.g.~the current verbosity level
becomes cumbersome, if the corresponding \code{write} statements are cluttered
throughout the code. Obviously the standard idea to reduce code duplication
also applies here, i.e.~one delegates the actual I/O operations, including
aspects like indentation, verbosity checking, etc. to a designated part of
the code, let's say a \code{log\_manager} module.

An additional difficulty arises in MPI-parallel applications. Typically only
one MPI process e.g.~the one with rank 0, is intended to generate log messages.
This implies that any generation of a log message must be wrapped inside code
for checking the rank of the executing MPI process. Again this check could
conveniently be encapsulated in said \code{log\_manager} module, thus,
uncluttering the rest of the code.

Naturally the delegation of these decisions and code parts to a separate module
comes at a price. However, performance-wise the extra costs resulting from
calls to a different subprogram in such an approach can be considered
uncritical as in a typical simulation code the time spent for generating log
messages is negligible compared to the actual computational work.

While our hypothetical \code{log\_manager} module handles the actual I/O
operations and possibly also deals with formatting issues and the like,
the generation of the actual text of a log message necessarily must happen
in the respective code part of the caller. The resulting text string must then
be send to the \code{log\_manager} module e.g.~by invoking a
\code{log\_write()} subroutine.

Generating a string literal to pass to a subroutine is of course
straightforward and not different in Fortran than in any other
programming language employed in scientific programming
\begin{codelet}
\begin{Verbatim}[frame=single,numbers=none]
call log_write( "Starting assembly of FE matrix" )
\end{Verbatim}
\end{codelet}
However, often log messages will contain information that is only available at
run-time, such as the number of vertices in an input mesh, the norm of a
residual vector or the current iteration count of a loop. Thus, we need to be
able to \emph{dynamically} generate messages. Of course, this is nothing one
could not accomplish using standard Fortran tools. We just need to perform an IO
operation on an ``internal file'', i.e.~an existing string. Assume that for
this purpose our \code{log\_manager} module provides a string of constant
length
\begin{codelet}
\begin{Verbatim}[frame=single,numbers=none]
character(len=max_message_length) :: msg
\end{Verbatim}
\end{codelet}
Then logging e.g.~the norm of a residual vector stored in the variable
\code{resnorm} could be achieved by
\begin{codelet}
\begin{Verbatim}[frame=single,numbers=none]
write( msg, "( 'resnorm = ', E12.4E3)" ) resnorm
call log_write( msg )
\end{Verbatim}
\end{codelet}
%
%
Or we could output the entry  $a_{2,3}$ of a matrix via
\begin{codelet}
\begin{Verbatim}[frame=single,numbers=none]
write( msg, "('(',I0,',',I0,') = ',F0.3)" ) 2, 3, a(2,3)
call log_write( msg )
\end{Verbatim}
\end{codelet}
giving us an output of e.g.~\verb*!(2,3) = 12.345!, where we used the F2008
feature of letting the program determine the field width automatically.

It would be nice, if one could reduce the amount of code by combining the
two lines into one, while at the same time getting rid of all the format
conversion specifiers. The latter could, of course, be accomplished with
list-directed I/O. But in this case the results would be compiler-dependent,
and often not very pleasing.

In this note we are going to present an approach to dynamic generation of
strings in Fortran that combines a high-flexibilty with a very low programming
effort at the place where it is employed. Syntactically the approach has a
look-and-feel similar to the stream--oriented I/O syntax in C++.

In the following we will describe it and provide several examples. A fully
working demonstrator implementation is available at \cite{Code:on:Zenodo}.

\section{General Concept}
\label{sec:GeneralConcept}

The general concept of our proposed approach to dynamic string generation is
straightforward. We combine the following well known ingredients
\begin{itemize}
\item Fortran's string concatenation operator \code{//}
\item deferred-length strings
\item overloading of functions and operators
\item optional arguments
\end{itemize}
which are described in detail in
e.g.~\cite{Chivers:2015:Book,Metcalf:2023:Book}.

One can see the approach as being composed of two layers. The first one is
functionality to convert non-character values into strings and the second
one extends it to its stream-like form.

\subsection{Stringification}
\label{subsec:GeneralConcept}
The central component of the first layer is a \code{v2s()} function. It handles
the conversion of a non-character value, e.g.~an integer or a floating point
number, into an (external) character representation. We denote this as
\emph{stringification}\footnote{We borrow this nomenclature for
 string-conversion from the stringification operator \# of the C
preprocessor, that allows to convert the argument of a parameterised macro
to a string constant.}. The name is an easy mnemonic: \emph{value-to-string}.
Before going into the technical details we provide three simple examples
of the use of the function

\begin{codelet}
\begin{Verbatim}[frame=single,numbers=none]
call log_write( 'iteration count = ' // v2s(iter) )
call log_write( 'eps  = ' // v2s(eps,'F8.2') )
call log_write( 'flag use_petsc is ' // v2s(.true., "switch" ) )
\end{Verbatim}
\end{codelet}

In each case \code{v2s()} returns a string that we concatenate on-the-fly
with the string literal on the left. 
Different datatypes, of course, require different conversion functions. Thus,
\code{v2s()} needs to be an overload. Assuming that the three types above
are \code{integer}, \code{real} and \code{logical} we can use (in a module
named \code{stringify})

\begin{codelet}
\begin{Verbatim}[frame=single,numbers=none]
interface v2s
   module procedure io_int2str, io_real2str, io_bool2str
end interface
\end{Verbatim}
\end{codelet}

Let us examine the first example. The argument to the \code{log\_write()}
subroutine is generated by concatenation of two strings. The second of these
is the return value from the call to \code{v2s(iter)}. As \code{iter} is assumed
to by a 32-bit integer the actual function to be invoked is \code{io\_int2str}.
Its source code is given in Fig.~\ref{fig:io_int2str}. As can be seen, it
expects two input arguments, \code{val} the integer value to stringify and
an optional \code{spec} argument. Line 5 shows that the function will
return a character string of deferred-length, which means that the program
itself will handle (re)allocation of it. As this only works on assignments,
but not with \code{write}, the actual write statement in line 17 makes
use of a \code{buffer} string of fixed length defined in the \code{stringify}
module. Generation of the return string happens in line 18, where we also
remove trailing whitespace with the help of \code{trim()}.

\begin{figure}[!b]
\begin{codelet}
\begin{Verbatim}[frame=single,numbers=left]
function io_int2str( val, spec ) result(str)

  integer(int32),             intent(in) :: val
  character(len=*), optional, intent(in) :: spec
  character(len=:), allocatable          :: str

  character(len=1024) :: fmt

  ! if conversion specifier present: use it
  if ( present(spec) ) then
     write( fmt,"(3A)") "(", spec, ")"
     write( buffer, fmt ) val
  else
     fmt = "(I0)"
  end if

  write( buffer, fmt ) val
  str = trim( buffer )

end function io_int2str
\end{Verbatim}
\end{codelet}
\caption{Example of stringification function for 32-bit integers.
\label{fig:io_int2str}}
\end{figure}
The second example is different in two respects. This time the argument to
\code{v2s()} is assumed to be of type \code{real}, with kind \code{real64},
and we provide our own conversion specification as a second argument. The
latter is not required, as the approach allows us to define, as part of the
\code{stringify} module, (project-wide) rules for the format conversion. An
example can be found in the demonstrator code, \cite{Code:on:Zenodo}.

On the other hand, providing a conversion specification also e.g.~allows us
to change the way logicals get converted. Our example \code{io\_bool2str()}
implementation provides the options \code{default} (classic Fortran-style, 
i.e.~T/F),
\code{word} (true/false), \code{code} (.true./.false.) and "switch" (on/off).
Thus, the third example passes \verb*!flag use petsc is on! to
\code{log\_write()}.
\subsection{C++-style Output}
\label{subsec:streamstyle}
Based on the stringification functionality introduced above we can now very
easily add a second conceptual layer that allows us to generate strings in
a C++-like fashion
\begin{codelet}
\begin{Verbatim}[frame=single,numbers=left]
call log_write( "Residual after " // it_count // &
       " iterations is " // res_norm )
\end{Verbatim}
\end{codelet}
For this we simply need to overload Fortran's concatenation operator. We
separate this layer into its own \code{streamstyle} module.
\begin{codelet}
\begin{Verbatim}[frame=single,numbers=left]
interface operator(//)
   module procedure int2stream, real2stream, bool2stream
end interface operator(//)
\end{Verbatim}
\end{codelet}
As \code{//} is a binary operator, the individual functions in the interface
all require two arguments. Since concatenation happens left-to-right, the
first one needs to be of string-type, while the second one will be of the
data-type to be converted. Leaving the specifics of the conversion to the
respective overload of \code{v2s()} makes the functions here syntactically
completely identical. As an example we present the one for \code{real64}.
\begin{codelet}
\begin{Verbatim}[frame=single,numbers=left]
function real2stream( str_in, val ) result( str_out )

  character(len=:), allocatable :: str_out
  character(len=*), intent(in)  :: str_in
  real(real64)    , intent(in)  :: val

  str_out = str_in // v2s( val )

end function real2stream
\end{Verbatim}
\end{codelet}

\section{Extensions}
\label{sec:Extensions}

\subsection{Multi-Line Messages}
%
Our first extension will allow generation of mult-line messages. Classically
one would handle this in an output statement in Fortran by inserting an
end-of-record specifier '\code{/}' into the conversion format. So
\begin{codelet}
\begin{Verbatim}[frame=single,numbers=none]
print "(A,/,A)", "1st line of message ...", "... and 2nd line"
\end{Verbatim}
\end{codelet}
will produce
\begin{verbatim*}
1st line of message ...
... and 2nd line
\end{verbatim*}
A simple alternative is to directly insert the line-break into the string.
For this purpose the \code{stringify} module provides a string literal
\code{newline} that is a shorthand for the ASCII line-feed control
character\footnote{This works for Unix/Linux and Mac OS. In the case of the
Windows OS we need to replace this by a carriage return followed by a
line-feed.}.
\begin{codelet}
\begin{Verbatim}[frame=single,numbers=none]
character(len=*), parameter :: newline = char(10) ! LF only
\end{Verbatim}
\end{codelet}
Hence, the following codelet
\begin{codelet}
\begin{Verbatim}[frame=single,numbers=none]
print "(A)", newline// " We can" // newline // " use multi" // &
     "-line" // newline // " strings, too!"
\end{Verbatim}
\end{codelet}
will output
\begin{verbatim*}
 We can
 use multi-line
 strings, too!
\end{verbatim*}
The same can also be done with tab-stops.

\subsection{I/O for Arrays}

The approach is not restricted to scalar types, but can be
extended to any kind of derived type. We start by giving an example for 
rank-1 arrays of integers, before considering a user-defined type.

We implement conversion of a rank-1 integer array in the function
\code{io\_intvec2str()} given in Fig.~\ref{fig:io_intvec2str}. Assume that
the variable \code{int\_vec} has the following four entries $(1,7,-3,5)$,
then the codelet
\begin{codelet}
\begin{Verbatim}[frame=single,numbers=none]
print "(A)", "An integer vector:" // newline // v2s(int_vec, "I2")
\end{Verbatim}
\end{codelet}
will print
\begin{verbatim}
An integer vector:
|  1 |
|  7 |
| -3 |
|  5 |
\end{verbatim}
\begin{figure}[b]
\begin{codelet}
\begin{Verbatim}[frame=single,numbers=left]
function io_intvec2str( val, spec ) result(str)

  integer(int32), dimension(:), intent(in) :: val
  character(len=*), optional               :: spec
  character(len=:), allocatable            :: str

  integer :: idx

  str = "| " // v2s(val(1),spec) // " |"
  do idx = 2, size(val)
     str = str // newline
     str = str // "| " // v2s(val(idx),spec) // " |"
  end do

end function io_intvec2str
\end{Verbatim}
\end{codelet}
\caption{One possible way to convert a rank-1 array of integers into a
string representation.\label{fig:io_intvec2str}}
\end{figure}
%
\subsection{Example with a User-Defined Structure Type}
%
We now demonstrate the extension to user-defined derived types. As an example
let use consider a simple type \code{point3d\_t} for representing a point in
3D.

The first step is to add a type-bound procedure \code{stringify\_point}
(lines 3-4)
\begin{codelet}
\begin{Verbatim}[frame=single,numbers=left]
type :: point3d_t
   real(real64) :: x, y, z
 contains
   procedure, pass (this) :: stringify_point
end type point3d_t
\end{Verbatim}
\end{codelet}
The procedure then might be implemented as follows
\begin{codelet}
\begin{Verbatim}[frame=single,numbers=left]
function stringify_point( this ) result( str_out )
  class(point3d_t), intent(in)  :: this
  character(len=:), allocatable :: str_out

  ! allocate deferred length string for using it in write()
  integer :: str_out_len = 6 + 3*4
  allocate( character(len=str_out_len) :: str_out )

  ! pretty-print point coordinates as triple
  write( str_out, "(3(A,SP,F3.1),A)" ) "(", this%x, ", ", &
     this%y, ", ", this%z, ")"
end function stringify_point
\end{Verbatim}
\end{codelet}
As second step we add a corresponding function to our \code{stringify} module
whose only task is to delegate string generation to the type-bound procedure.
\begin{codelet}
\begin{Verbatim}[frame=single,numbers=left]
function io_point3d2str( val ) result(str)
  class(point3D_t), intent(in)  :: val
  character(len=:), allocatable :: str
  str = val%stringify_point()
end function io_point3d2str
\end{Verbatim}
\end{codelet}
This new function \code{io\_point3d2str()} is then added as another possiblity
to the overload list for \code{v2s()}. In order to simply place objects of
type \code{point3d\_t} in the \enquote{output stream}, we extend as third and
final step the \code{streamstyle} module. For this we add a function
\code{point2stream()}, which, again, is syntactically identical to
\code{real2stream()} above, and add it to the list of overloads for the
concatenation operator.
\begin{codelet}
\begin{Verbatim}[frame=single,numbers=left]
interface operator(//)
   module procedure int2stream, real2stream, point2stream
end interface operator(//)
\end{Verbatim}
\end{codelet}
Now the following codelet
\begin{codelet}
\begin{Verbatim}[frame=single,numbers=left]
type(point3d_t) :: point = point3d_t( 0.5, 1.0, -2.0 )
write(*,'(A)') "Point at coords = " // point // " in domain"
\end{Verbatim}
\end{codelet}
will output \verb*!Point at coords = (+0.5, +1.0, -2.0) in domain!.
%
\subsection{Manipulators}
%
In the stream-like approach of Sec.~\ref{subsec:streamstyle} we just insert
data-objects into the \enquote{stream}, but have no direct way to influence
their conversion to a character representation. C++ solves this issue by
allowing to insert special manipulator objects into the stream that steer the
conversion of items following afterwards (and additional stream properties).
For a quick overview, see e.g.~\cite{CPP:IO:MANIP}.

In our setting we could address this issue, by not passing the data-items
to the \enquote{stream} directly, but first feeding them through \code{v2s()},
while providing the desired specification. However, in order to demonstrate
the versality of our approach, we will now show how to emulate the concept of
I/O manipulators with an example.

The standard C++ library defines two objects \code{std::showpos} and
\code{std::noshowpos}. The former one activates display of a plus sign for
positive numbers put into the stream, while the latter deactivates it. We
can achieve the same by adding to the \code{stringify} module

\begin{itemize}
\item a logical variable \code{show\_sign}
\item two functions \code{showpos} and \code{noshowpos}
\item and including \code{show\_sign} into our default formatting rules
\end{itemize}

Below we give the source code for the \code{showpos} function. The one for
\code{noshowpos} is identical, apart from setting the flag to
\code{.false.}$\,$. The function returns an empy string, in order to be
compatible with the overloading of \code{//} and to not visibly change the
string assembly.
\begin{codelet}
\begin{Verbatim}[frame=single,numbers=left]
function showpos() result( nochar )
  character(len=:), allocatable :: nochar
  nochar = ""
  show_sign = .true.
end function showpos
\end{Verbatim}
\end{codelet}
In a similar fashion other manipulators, e.g.~\code{setprecision} (for
changing floating-point conversion) or \code{setw} (for setting the field
width in the conversion) could be implemented.
%
%
\section{Portability}
\label{sec:PortabilityIssues}
All the components used in our approach conform to the Fortran standard. One
minimally needs F2008, as this includes automatic field-width determination
(\code{I0,F0.*}, \ldots) and nested I/O operations (first allowed in F2023). As
these
are well-implemented, see e.g.~\cite{Chivers:2019:ACMFF}, there should be no
portability issues. The demonstrator at \cite{Code:on:Zenodo} has successfully
been test with the following compilers:
\begin{center}
\begin{tabular}{lll}
\hline\noalign{\smallskip}
\multicolumn{2}{c}{Compiler} & version \\
\noalign{\smallskip}\hline\noalign{\smallskip}
AMD/AOCC      & flang     & 12.0.0    \\
GNU           & gfortran  & 12.2.0    \\
Intel/classic & ifort     & 2021.11.1 \\
Intel/OneAPI  & ifx       & 2024.0.2  \\
Nvidia        & nvfortran & 21.9-0    \\
\noalign{\smallskip}\hline
\end{tabular}
\end{center}
%
%
\bibliographystyle{ACM-Reference-Format}
\bibliography{references}
%
%
\end{document}